\documentclass[3p,times,twocolumn]{elsarticle}
\usepackage{epsfig}
\usepackage{xcolor}
\usepackage{epsfig,graphicx,float}
\usepackage{mathrsfs}
\usepackage[utf8]{inputenc}
\usepackage{url}
\usepackage[caption = false]{subfig}
\usepackage{amsmath}
\usepackage[normalem]{ulem}
\usepackage{multirow}
\usepackage{braket}
\usepackage{bm}
\begin{document}
	\newcommand{\gs}[1]{\textcolor{red}{\it #1}}
	\newcommand{\gsout}[1]{\textcolor{red}{\sout{#1}}}
	\newcommand{\rc}[1]{\textcolor{blue}{\it #1}}
	\newcommand{\rcout}[1]{\textcolor{blue}{\sout{#1}}}

	\begin{frontmatter}
	
		\title{The $^{19}$N($n$,$\gamma)^{20}$N capture rate in light of the probable bubble nature of $^{20}$N}
		
		\author[1]{V. Choudhary}
		\ead{vchoudhary@ph.iitr.ac.in}
		
		\author[2]{M. Dan}
		\ead{mdan.phy@nitrr.ac.in}
		
		\author[1]{R. Chatterjee}
		\ead{rchatterjee@ph.iitr.ac.in}
		
		\author[3,4]{M. Kimura}
		\ead{masaaki.kimura@riken.jp}
		
		\author[5,6,3,7]{W. Horiuchi}
		\ead{whoriuchi@omu.ac.jp}
		
		\author[8]{Shubhchintak}
		\ead{shubhchintak@ulb.ac.be}

		\author[9,10]{G. Singh}
		\ead{g.singh@unipd.it}

		\address[1]{Department of Physics, Indian Institute of Technology Roorkee, Roorkee 247 667, India}
		\address[2]{Department of Physics, National Institute of Technology Raipur, Raipur, 492010, India.}
		\address[3]{RIKEN Nishina Center, Wako 351-0198, Japan}
		\address[4]{Nuclear Reaction Data Centre, Faculty of Science, Hokkaido University,
			060-0810 Sapporo,Japan}
		\address[5]{Department of Physics, Osaka Metropolitan University, Osaka 558-8585, Japan}
		\address[6]{Nambu Yoichiro Institute of Theoretical and Experimental Physics (NITEP), Osaka Metropolitan University, Osaka 558-8585, Japan  }
		\address[7]{Department of Physics, Hokkaido University, Sapporo 060-0810, Japan}
		\address[8]{Physique Nucl{\'{e}}aire Th{\'{e}}orique et Physique Math{\'{e}}matique, C.P. 229, Universit{\'{e}} Libre de Bruxelles (ULB), B-1050 Brussels, Belgium}
		\address[9]{Dipartimento di Fisica e Astronomia ``G.Galilei'', Università degli Studi di Padova, via Marzolo 8, Padova, I-35131, Italy}
		\address[10]{INFN-Sezione di Padova, via Marzolo 8, Padova, I-35131, Italy}

		\date{\today}

		\begin{abstract}
	We aim to explore the bubble nature of the exotic nucleus $^{20}$N within the microscopic antisymmetrized molecular dynamics (AMD) approach. Constraining its structural parameters, we  analyse its static properties. Subsequently, we use the AMD infused finite-range distorted-wave Born approximation theory to calculate the Coulomb breakup of $^{20}$N as an indirect approach to estimate the $^{19}$N$(n,\gamma)^{20}$N radiative capture rate.

		\end{abstract}

	\end{frontmatter}

	\section {Introduction}
	
Nuclei away from the valley of stability can exhibit exotic behaviors such as the halo phenomenon
\cite{TSK13PPNP}, but movement along this coast towards the medium and heavy mass region also
results in the manifestation of other such peculiar properties. One such feature evolves to produce
a rearrangement of the density profile of a nucleus, decreasing it right at the center. These atomic
cores with slightly decreased nuclear densities near the origin are called ``bubble'' nuclei
\cite{MLS17NatP,CHK20PRC,LLS16PRC}. Such bubble-like formations in non-cluster nuclei are unexpected
as they challenge the quantum liquid picture of a typical nucleus \cite{MLS17NatP,EKN12Nat}. 

The central depletions in the nuclear density usually are a delicate result of a reduced occupation of
single-particle orbits with low orbital angular momentum, whose wave functions {peak mainly in} the
nuclear interior. On the other hand, {the presence of higher angular momentum components in the
nuclear wave functions suppresses} the central density \cite{MLS17NatP, GGK09PRC}. An implication of
this phenomenon is that bubble nuclei usually do not have an $s$-state contribution 
and are mostly populated by the orbits with larger angular momentum. It has been
shown that such nuclei are usually sharper in their nuclear 
surface region \cite{CHK20PRC}, \textit{i.e.,} the nuclear diffuseness is very small, with the
differences in the diffuseness for the bubble and non-bubble nuclei being quite prominent. In a
sense, this implies a relation between the occupation of orbital angular momentum states by nucleons
and the nuclear surface diffuseness~\cite{CHK20PRC,HHK18PRC,H21PTEP,CHK21PRC,horiuchi2022density,horiuchi2023}. Nevertheless, the extraction of the bubble information on the nuclear profile is challenging as the applications of electron scattering
measurements~\cite{TEO17PRL} for unstable nuclei is still limited. Thus, various reaction observables
should be investigated in this regard to decipher the matter density distributions of nuclei. 

Extensive studies have been carried out for established as well as potential bubble candidates in
the medium mass region, like $^{34}$Si \cite{MLS17NatP,GGK09PRC}, $^{46}$Ar \cite{NSM13PRC} and
$^{22}$O \cite{CHK20PRC} and some in the heavy and superheavy region
\cite{SNR17PRC,BRR99PRC,HHK18PRC,PA22PRC}. We, for the present work, concentrate on a Nitrogen isotope near
the drip line, $^{20}$N. In a simple shell model picture, the outermost 5 neutrons occupy the
$d_{5/2}$ orbit. Hence, the bubble structure is expected in the ground state of $^{20}$N, much like
$^{22}$O. Apart from its exotic character, we select this nucleus due to its astrophysical relevance
as it is supposed to be crucial to the final abundance of Fluorine isotopes in the neutrino-wind
driven model of $r$-process nucleosynthesis \cite{TSK01AJ,CD20N}. 

 Given these motivations, we study the $^{19}$N($n$,$\gamma)^{20}$N radiative capture reaction
 through the indirect method of Coulomb dissociation (CD). Recently, the method of CD was applied to
 an experimental study of $^{20}$N \cite{CD20N}, where a secondary radioactive ion beam of $^{20}$N was  made to impinge on a $^{208}$Pb target and dissociate into a $^{19}$N
 core and a neutron. Inverse 
 kinematics was then applied to study the radiative neutron capture reaction
 $^{19}$N($n$,$\gamma)^{20}$N. Theoretically, the method of CD has been reliably used in the past
 for studying several capture reactions of astrophysical interest
 \cite{BCS08PRC,SSC17PRC,BB88PR,SCS17PRC,DSC19PRC} and is one of the many available indirect tools
 available to study capture processes at the relevant energy regime
 \cite{BB88PR,CRA16PRC,TBC14RPP,BERTULANI2010195}. 

We apply CD through the fully quantum mechanical post-form  finite-range distorted-wave Born approximation (FRDWBA) theory \cite{RCNPA,CS18PPNP,SSC16PRC}, whose only input is the ground state (g.s.) wave function. We use this breakup theory to compute the relative energy spectra. Subsequently, we invoke the principle of detailed balance to calculate the radiative capture cross section from the photodisintegration cross section, which is obtained via the elastic dissociation of $^{20}$N under the dynamic Coulomb field of $^{208}$Pb. Finally, from the radiative capture cross section, we derive the total reaction rate for the $^{19}$N($n$,$\gamma)^{20}$N radiative capture reaction. We compare our results obtained using the microscopic structure inputs from the antisymmetrized molecular dynamics (AMD) method \cite{kanada2003, kanada2012, kimura2016}, as well as AMD with a ``tail-corrected'' approach \cite{DCK21EPJ}, and also with a phenomenological method.

In the following section, a brief description of the formalism is presented. An assessment of the results obtained from our calculations is done in section \ref{rnd}, followed by a conclusion.

\section{Theoretical formalism}

\subsection{ Finite-range distorted-wave Born approximation (FRDWBA)}
The triple differential cross section for \textit{d} + \textit{t} $\longrightarrow$ \textit{b} + \textit{c} + \textit{t} reaction is written as 
\begin{equation}
		\dfrac{d^3\sigma}{dE_{b}d\Omega_{b}d\Omega_{c}} = \dfrac{2\pi}{\hbar v_{dt}}\rho{(E_{b},\Omega_{b},\Omega_{c})}\sum_{l,m}|\beta_{lm}|^{2}.
		\label{a2.1}
\end{equation}
 In our case, $d$ is the $^{20}$N projectile, $t$ is the $^{208}$Pb target, $b$ is the $^{19}$N core, and $c$ represents the outgoing neutron ($n$). The phase space factor of the three-body final state is represented by $\rho{(E_{b},\Omega_{b},\Omega_{c})}$ \cite{Fuchs}.  
$v_{dt}$ is the relative velocity between the projectile and target in the initial channel. 
$\beta_{lm}$ is the reduced transition amplitude that incorporates the ground state radial wave function, $u_l(r)$, of the projectile having any angular momentum $l$, and its projection $m$. While the best case scenario for this input is one derived from a microscopic many body method like the AMD \cite{DCK21EPJ}, oftentimes a phenomenological approach has been used. 

Once we have the triple differential cross section from the FRDWBA theory, we can easily obtain the relative energy spectrum through appropriate integration and multiplication with the appropriate Jacobian \cite{BCS08PRC,CS18PPNP}.  
For a dipole dominated pure Coulomb breakup, the relative energy spectrum $({d\sigma}/{dE_{rel}})$ can be related to the photodisintegration cross section, $\sigma_{(\gamma,n)}$, as \cite{BBR86NPA}
\begin{equation}
	\begin{aligned}
		\dfrac{d\sigma}{dE_{rel}} = \dfrac{1}{E_{\gamma}}{\sigma_{(\gamma,n)}}{n_{E1}},
	\end{aligned}
	\label{relt.eq}
\end{equation}
where $n_{E1}$ is the virtual photon number. The $\gamma$-energy,
$E_{\gamma}= E_{rel} + S_n$, where $S_n$ is the one neutron separation energy, and $E_{rel}$ is the relative energy between fragments \textit{b}-\textit{c} in the final channel. Knowing the photodisintegration cross section, the time reversed radiative capture cross section, $\sigma_{(n,\gamma)}$, can be computed by using the principle of detailed balance \cite{BBR86NPA} as

\begin{equation}
\begin{aligned}
\sigma_{(n,\gamma)} = \dfrac{2(2\textit{j}_d+1)}{(2\textit{j}_b+1)(2\textit{j}_c+1)}\dfrac{\textit{k}_\gamma^2}{\textit{k}^2}{\sigma_{(\gamma,n)}},
\end{aligned}
\label{balance.eq}
\end{equation}
where, $j_b$, $j_c$, and $j_d$ are the total spins of the particle $b$, $c$, and $d$, respectively. $\textit{k}_\gamma$ is the photon wave number and $\textit{k} = \dfrac{2\mu_{bc} E_{rel}}{\hbar^2}$, with $\mu_{bc}$ being the reduced mass of the \textit{b}-\textit{c} system.

Finally, the reaction rate per mole for the radiative capture reaction $(\textit{b} + \textit{c}
\longrightarrow  \textit{d} + \gamma )$ is $R=N_A \langle\sigma_{(n,\gamma)}\textit{v}\rangle$,
where, $N_A$ is the Avogadro number and $\langle\sigma_{(n,\gamma)}\textit{v}\rangle$ is the
reaction rate per particle pair and is averaged over the Maxwell-Boltzmann velocity distribution
\cite{Rolfs}. 
For more details one is referred to \cite{BCS08PRC,SSC17PRC,CS18PPNP,NSC15PRC}.

\subsection{Antisymmetrized molecular dynamics (AMD) method}
In this paper, we employ the AMD method to describe the ground states of $^{19}{\rm N}$ and $^{20}{\rm N}$. The Hamiltonian is given by
\begin{align}
H = \sum_{i=1}^A t_i + \sum_{i<j}^A v_{ij} - t_{\rm cm},
\end{align}
which is composed of the kinetic energies of nucleons and the Gogny D1S density
functional~\cite{berger1991} which includes an effective nucleon-nucleon interaction and Coulomb
interactions. The center-of-mass kinetic energy, $t_{\rm cm}$, is exactly removed. The variational wave function of AMD is a parity-projected Slater determinant of
single-particle Gaussian wave packets
\begin{align}
 \Phi^\pi_{\rm int} &=\frac{1+\pi P_x}{2}
 {\mathcal A} \set{\varphi_1\varphi_2\cdots\varphi_A},  \label{eq:intwf}  
\end{align}
where $P_x$ is the parity operator. In this study, we only calculate the negative-parity states
since we are interested in the low-lying states. 
A single-particle wave packet is represented by a deformed Gaussian~\cite{kimura2004}
\begin{align}
 \varphi_i({\bf r}) &= \prod_{\sigma=x,y,z}
 \exp\set{-\nu_\sigma(r_\sigma -Z_{i\sigma})^2}\chi_i\eta_i,
 \label{eq:singlewf} 
\end{align}
where $\chi_i$ is the spinor, and $\eta_i$ is the isospin fixed to either a proton or neutron.

The parameters of the variational wave function are ${\bm Z}_i$, $\bm \nu$, and $\chi_i$, which were
determined by the energy variation with the constraint on the nuclear quadrupole deformation
parameter $\bar{\beta}$. No constraint is imposed on the other quadrupole deformation parameter $\bar{\gamma}$. 
Therefore, its value is determined to be the optimal value for each given $\bar{\beta}$, which minimizes the energy. By the energy variation, we obtained the wave functions $\Phi^{\pi}_{\rm
int}(\bar{\beta})$ which have the minimum energy for  each given value of $\bar{\beta}$. They were projected on
the eigenstates of the total angular momentum
\begin{align}
 \Phi^{J\pi}_{MK}(\bar{\beta}) &= P^{J}_{MK}\Phi^{\pi}_{\rm int}(\bar{\beta}) \nonumber \\
 &=\frac{2J+1}{8\pi^2} \int d\Omega D^{J*}_{MK}(\Omega) R(\Omega)\Phi^{\pi}_{\rm int}(\bar{\beta}),
 \label{eq:prjwf} 
\end{align} 
where $P^{J}_{MK}$, $D^{J}_{MK}(\Omega)$, and ${R}(\Omega)$ denote the angular momentum projector,
Wigner $D$ function, and rotation operator, respectively. The wave functions with different values
of the quadrupole deformation parameter $\bar{\beta}$ are superposed to describe the ground and excited states
\begin{align}
 \Psi^{J\pi}_{M\alpha} = \sum_{Ki} e_{Ki\alpha}\Phi^{J\pi}_{MK}(\bar{\beta}_i).\label{eq:gcmwf} 
\end{align}
The coefficients $e_{Ki\alpha}$ can be obtained from the Hill-Wheeler equation~\cite{hill1953}
\begin{align}
 &\sum_{K'i'}(H_{KiK'i'}-E_\alpha N_{KiK'i'})e_{K'i'\alpha} = 0,\\
 &H_{KiK'i'} = \braket{\Phi^{J\pi}_{MK}(\bar{\beta}_i)|H|\Phi^{J\pi}_{MK'}(\bar{\beta}_{i'})}, \\
 &N_{KiK'i'} = \braket{\Phi^{J\pi}_{MK}(\bar{\beta}_i)|\Phi^{J\pi}_{MK'}(\bar{\beta}_{i'})},
\end{align}
where $E_\alpha$ is the eigenenergy of the eigenfunction given by Eq.~(\ref{eq:gcmwf}).

The valence neutron wave functions are calculated from the overlap of the AMD wave functions for
$A-1$ and $A$ nuclei \cite{kimura2017}, {\it i.e.} $^{19}{\rm N}$ and $^{20}{\rm N}$  
\begin{align}
 \psi(\bm r) = \sqrt{20}\braket{\Psi^{1/2^-}_{-1/2}(^{19}{\rm N})|
 \Psi^{2^-}_{0}(^{20}{\rm N})}, \label{eq:ofunc1}
\end{align}
where bra and ket are the wave functions for the ground states of $^{19}{\rm N}$ ($J^\pi=1/2^-$) and $^{20}{\rm N}$ ($J^\pi=2^-$), respectively. Then, we calculate the multipole
decomposition of $\psi(\bm r)$
\begin{align}
 \psi(\bm r)=\sum_{jl}C^{20}_{1/2\,-1/2,j\,-1/2}\frac{u_{jl}(r)}{r}[Y_l(\hat r)\otimes \chi]_{jM},\label{eq:ofunc2}
\end{align}
where $C^{20}_{1/2\,-1/2,j\,-1/2}$ is the Clebsch-Gordan coefficient.  $u_{jl}(r)$ is the valence neutron wave
function in the $J^\pi\otimes l_j$ channel where the core nucleus 
$^{19}{\rm N}$ with spin-parity $J^\pi$ is coupled to the valence neutron with total angular
momentum $j$ and orbital angular momentum $l$. Its square-integral is the spectroscopic factor 
\begin{align}
 C^{2}S = \int_0^\infty dr\ |u_{jl}(r)|^2.
\end{align}
A problem is that $u_{jl}(r)$ obtained from the AMD wave functions does not have the correct
asymptotic behavior at large distances because the single-particle wave packet
is limited to a Gaussian form. Hence, we introduce the modified neutron wave function
$\tilde{u}_{jl}(r)$ which is identical to $u_{jl}(r)$ in the internal region and smoothly connected
to the exact 
asymptotic form at the channel radius $r=a$
\begin{align}
 \tilde{u}_{jl}(r)=
  \begin{cases}
   u_{jl}(r) & \text{if $r\leq a$,} \\
   C\kappa rk_l(\kappa r)       & \text{if $r\geq a$,}
  \end{cases}
\end{align}
where $k_l$ is the modified spherical Bessel function of second kind, and a factor
$\kappa=\sqrt{2\mu S_n/\hbar^{2}}$ is determined by the reduced mass $\mu$ of the $n+{}^{19}{\rm N}$ system
and one-neutron separation energy $S_n$. The asymptotic normalization coefficient $C$ and the channel
radius $a$ are determined so that $u_{jl}(r)$ is smoothly connected to $C\kappa r k_l(\kappa r)$ 
at $r=a$.
\begin{align}
 \left. \frac{\rm d}{\rm dr} \ln u_{jl}(r)\right|_{r=a} &=  
 \left. \frac{\rm d}{\rm dr} \ln  rk_{l}(\kappa r)\right|_{r=a},\\
 u_{jl}(a) &= C\kappa ak_{l}(\kappa a).
 \label{channel_rd.eq}
\end{align}
Thus-obtained $\tilde{u}_{jl}(r)$ is used as an input of the reaction calculation.

\section{Results and discussions}
\label{rnd}
\subsection{Properties of the ground states of $^{19}{\rm N}$ and $^{20}{\rm N}$ calculated by AMD}
\begin{figure}[h!]
 \centering
 \includegraphics[width=0.8\hsize]{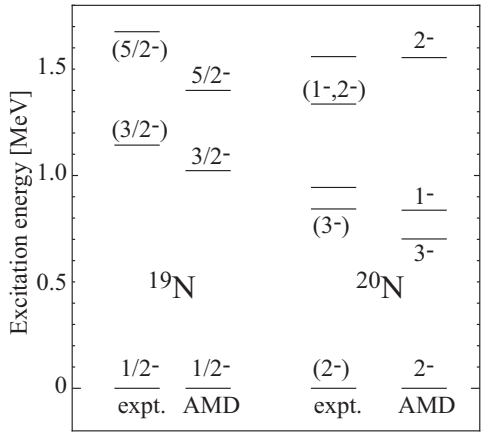}
 \caption{Calculated and observed low-lying spectra of $^{19}{\rm N}$ and $^{20}{\rm
 N}$.  The experimental data are taken from Ref.~\cite{NNDC}.} 
 \label{fig:level} 
\end{figure}

 Figure~\ref{fig:level} compares the excitation spectra of $^{19}{\rm N}$ and $^{20}{\rm N}$
 obtained by AMD with experimental data. The calculated spin-parity of the ground states are
 $1/2^-$ and $2^-$, respectively, in agreement with experimental data and shell-model
 calculations~\cite{NNDC,Yuan2012}. In-beam $\gamma$-ray spectroscopy experiments~\cite{Sohler2008,Elekes2010} 
 have observed several excited states for both nuclei, although their spin-parities have not been 
 firmly assigned. The ordering of the excited states of $^{19}{\rm N}$ is in agreement with
 the shell model calculations~\cite{Yuan2012}. On the other hand, for the excited state of 
 $^{20}{\rm N}$, the AMD and shell model calculations do not agree. Note that the
 ordering of the excited states changes depending on the effective interactions used in the shell
 model calculations~\cite{Yuan2012}. This suggests that the excitation spectrum of $^{20}{\rm N}$
 depends on details of the nucleon-nucleon interactions, such as spin-spin, tensor, and three-body
 forces.  
\begin{table*}
\begin{center}
 \caption{ Calculated ground-state properties of $^{19}{\rm N}$ and $^{20}{\rm N}$ compared with
 the experimental data~\cite{Kameda2004,OBC01NPA,bagchi2019}. $\bar{\beta}$ and $\bar{\gamma}$ denote matter quadrupole deformation
 parameters, where $\bar{\gamma}$ is given in degrees. The point proton and matter distribution radii
 ($R_p$ and $R_m$) are given in units of fm. The electric quadrupole moment and $g$-factor for the
 magnetic moment ($Q$ and $g$) are given in units of $\rm fm^2$ and dimensionless, respectively.}  
 \label{tab:gs}
\begin{tabular}{ccccccccccc}
 \hline
 &$J^\pi$ & $\bar{\beta}$ & $\bar{\gamma}$    & $Q$  & $g$ & $|g|$(expt.) & $R_p$ & $R_p$(expt.) & $R_m$  & $R_m$(expt.) \\
 \hline
 $^{19}{\rm N}$  & $1/2^-$   & 0.40 & 6  & --  & $-0.72$ & 0.61(3)~\cite{Kameda2004} & 2.66 & 2.52(3) &  2.85  & 2.74(3)~\cite{bagchi2019} \\
                 &         &      &    &     &         &         &      &       &         & 2.71(3)~\cite{OBC01NPA} \\
 $^{20}{\rm N}$  & $2^-  $ & 0.34 & 42 & 2.8 &  0.09   &         & 2.66 &  2.52(3) &2.87  & 2.84(5)~\cite{bagchi2019} \\
                 &         &      &    &     &         &         &      &       &         & 2.81(4)~\cite{OBC01NPA} \\
 \hline
\end{tabular}
\end{center}
\end{table*}

The properties of the ground states are summarized in Table~\ref{tab:gs}. Both nuclei are largely
deformed with the quadrupole deformation parameter $\bar{\beta}$ larger than 0.3, but their $\bar{\gamma}$ parameters are
different: $^{19}{\rm N}$ is prolately deformed, while $^{20}{\rm N}$ is oblately deformed. The electric
quadrupole moment and $g$-factor of magnetic moment are calculated without effective charge or
quenching factor. Experimentally, only the absolute value of the $g$-factor of $^{19}{\rm N}$ has
been reported~\cite{Kameda2004}, and the result of the AMD calculation is close to that. The AMD calculation
overestimates the observed proton and matter distribution radii except for the 
matter distribution radius of $^{20}{\rm N}$ \cite{OBC01NPA,bagchi2019}. This tendency was also found in other light
isotopes, such as Be and B isotopes. This is due to the strong repulsive density-dependent force of the
Gogny D1S density functional and the use of the common  Gaussian width parameter for protons and
neutrons in the present AMD framework.  

\subsection{Density distributions}
\label{3.2}  
We now turn our attention to the matter density distributions of $^{20}$N, calculated with the microscopic AMD and the phenomenological harmonic oscillator (HO) wave functions. The proton and neutron configurations of $^{20}$N for HO are  assumed to be $(0s)^{2}$$(0p)^{5}$ and $(0s)^{2}$$(0p)^{6}$$(0d)^{5}$, respectively. The valence neutron occupying the $d$-shell (without $1s$) renders a large enough orbital angular momentum to a nucleus to make it a probable bubble candidate \cite{CHK20PRC}. The size parameter ($\mu$ = 1.805 fm) is chosen to reproduce the root-mean-square (rms) radius of 2.84 fm \cite{bagchi2019}. 

\begin{figure}[!h]
	\centering
	\includegraphics[scale=0.45]{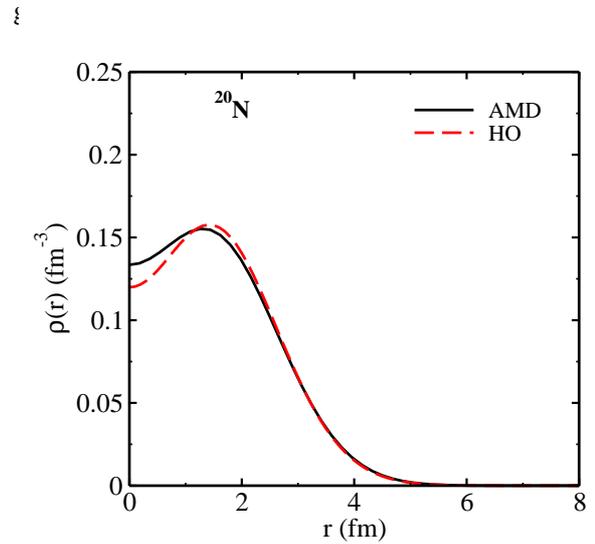}        
	\caption{ Matter density distributions of $^{20}$N obtained
		by AMD and the HO model.}
	\label{fig:density distribution}
\end{figure}

These density distributions (Fig.~\ref{fig:density distribution}) clearly depict a strong depletion of matter at the centre of the nucleus where the density is significantly lower than the maximum amplitude, which lies between 1.5 -- 2\,fm. Thus, it is highly probable that $^{20}$N is possibly shaped like a biconcave spheroid\footnote{In more common terms, like a human red blood cell.} and has a bubble character in its ground state.

The occupation probability near the Fermi surface reflects the density distributions near the nuclear surface~\cite{CHK20PRC,HHK18PRC,H21PTEP,CHK21PRC}. As the central density depression comes from the lack of the $s$-orbits, a bubble nucleus must exhibit a sharper nuclear surface than which includes $s$-orbits~\cite{CHK20PRC}.
Thus, it is logical to compute its diffuseness from the AMD and HO densities. For the same, we start from a phenomenological two-parameter Fermi (2pF) density distribution as
\begin{equation}
\rho_{F}(r) = \dfrac{\rho_{0}}{1+\exp\left[(r-R_{F})/a_{F}\right]}, 	
\end{equation}
where $a_{F}$ and $R_{F}$ are the diffuseness and radius parameters, respectively. The value of $\rho_{0}$ is uniquely determined for a given $a_{F}$ and $R_{F}$ by the normalization. We then determined the diffuseness parameter directly by minimizing the quantity~\cite{HHK18PRC}
\begin{equation}
\dfrac{4\pi}{A}\int_0^{\infty} |\rho(r)-\rho_{F}(r)|r^2 dr,
\label{minimize.eq}
\end{equation}
where $\rho$ is the density distribution obtained with AMD and the HO, and $A$ is the mass number of the nucleus. Using Eq.~(\ref{minimize.eq}), the values of diffuseness parameter for $^{20}$N are 0.51 fm for HO and 0.53 fm for the AMD (in the $d_{5/2}$-state). The nuclear diffuseness of HO is slightly lesser than AMD because the density of HO is more suppressed in the interior and more sharp around the nuclear radius than AMD. Nonetheless, the consistently lower diffuseness predicted by both the AMD and HO densities, seems to be a strong indication of the bubble structure for $^{20}$N~\cite{CHK20PRC}.

\subsection{Wave functions of the valence neutron}
Since we study $^{19}{\rm N}$($n$,$\gamma)^{20}{\rm N}$ and its inverse reaction,
$^{20}{\rm N}(\gamma,n)^{19}{\rm N}$, we are interested in the valence neutron wave function
coupled to the ground state of $^{19}{\rm N}$. The spin-parity of the ground states 
of $^{19}{\rm N}$ and $^{20}{\rm N}$ are $1/2^-$ and $2^-$, respectively. Hence, the valence neutron
must be in either $d_{5/2}$ or $d_{3/2}$ orbit. We have found several interesting features of the valence
neutron wave functions calculated by AMD as described below. The $C^{2}S$
(square-integral of $u_{jl}$) is 0.27 for the
$1/2^-\otimes d_{5/2}$ channel whereas that of the $1/2^-\otimes d_{3/2}$ channel is negligibly
small, 0.015. Thus, the valence 
neutrons mostly occupy $d_{5/2}$ orbit and rarely occupy the $d_{3/2}$ orbit. There is no
experimental data for the $C^{2}S$ to be compared with the AMD results. However, we note the
dominance of the $d_{5/2}$ orbit is consistent with the increase of the $N=14$ shell gap in
neutron-rich N isotopes reported by an experiment~\cite{bagchi2019}.

 We also consider a phenomenological description of the core-valence neutron interaction through a Woods-Saxon (WS) potential of the form 

\begin{equation} 
V(r) = V_{0} f(r) + V_{1} r_0^2\ell.s\dfrac{1}{r}\dfrac{d}{dr} f(r),
\end{equation}
where
\begin{equation}
f(r) = \bigg[1 + \exp\bigg(\dfrac{r-R_{\rm WS}}{a_{\rm WS}}\bigg)\bigg]^{-1}, \quad R_{\rm WS} = r_{0}A^{1/3}.
\end{equation}
Here, $V_0$ and $V_1$ are the potential depths of the central and spin-orbit potentials, respectively. $V_{1}$ is taken to be $\left[22-14(N-Z)/A  \right]$ \cite{BMBook}, and for our purpose, it comes out to be 17.8\,MeV. The radius and diffuseness
parameters are taken as $r_{0}$ = 1.25 fm and
$a_{\rm WS}$ = 0.65 fm, respectively. $V_{0}$ is then adjusted to reproduce the one-neutron separation energy of $^{20}$N ($S_{n}$ = $2.16$~MeV).

Note that the nuclear diffuseness is determined by the angular momentum of the single-particle orbits near the Fermi surface and strongly depends on the shell structure~\cite{H21PTEP}. As the $a_{\rm WS}$ includes the range of the nuclear force,
the value is in general larger than that extracted from the density distribution in Sec.~\ref{3.2}.

\begin{figure}[h!]
\centering
\includegraphics[scale=0.7]{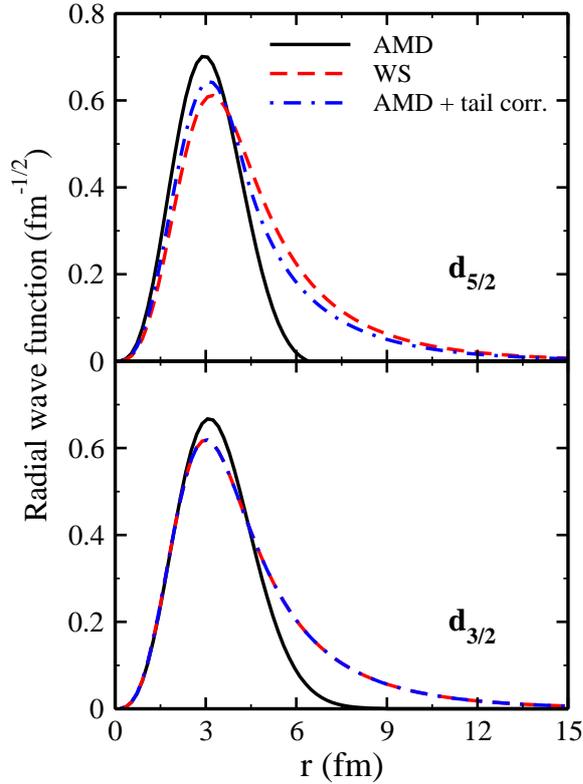}        
\caption{ Normalized wave functions of the AMD ($u_{jl}$), WS, and the AMD with tail correction ($\tilde{u}_{jl}$) models.}
\label{fig:wave_fn_all}
\end{figure}

With this description, we present in Fig. \ref{fig:wave_fn_all} a comparison of the valence neutron wave functions obtained from AMD and the phenomenological WS approaches.  
Comparing the normalized AMD ($u_{jl}$) and the AMD with tail correction ($\tilde{u}_{jl}$) shown in Fig.~\ref{fig:wave_fn_all}, it is clear that the original
AMD does not describe the correct asymptotic form and requires a correction at large distances.
The channel radii obtained from Eq.~(\ref{channel_rd.eq}) are 3.8 and 4.1 fm for $d_{5/2}$ and $d_{3/2}$,
respectively. Interestingly, a properly parameterized WS potential produces a valence neutron wave 
function quite similar to $\tilde{u}_{jl}$. It is noted that the correction of the asymptotic 
form changes the $C^{2}S$, but even after the correction, the $C^{2}S$ of $d_{3/2}$ orbit is
still two orders of magnitude smaller than that of $d_{5/2}$.  
 Therefore, when we discuss the radiative capture rates later on, we can safely assume that the valence neutrons occupy only the $d_{5/2}$ orbit.

\subsection{The reaction rates}
\label{sec: rates}

 We now consider the CD of $^{20}$N breaking elastically on $^{208}$Pb at 256 MeV/u beam energy. The beam energy was taken to be in consonance with the experimental conditions mentioned in Ref.~\cite{CD20N}. Subsequently, as outlined in Eqs.~(\ref{a2.1}),~(\ref{relt.eq}), and~(\ref{balance.eq}), we estimate the $^{19}$N($n$,$\gamma)^{20}$N radiative capture reaction rates.

\begin{figure}[h]
  \centering
  \includegraphics[scale=0.45]{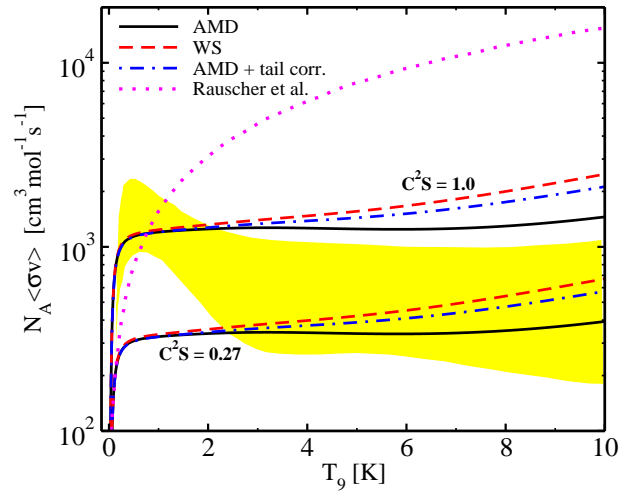}    \caption{ $^{19}$N($n$, $\gamma$)$^{20}$N reaction rates for the g.s. $d_{5/2}$ configuration of $^{20}$N. The AMD, WS and AMD with tail corrected results are with the AMD deduced $C^2S = 0.27$ for the $d_{5/2}$ state. The results with $C^2S = 1.0$, are shown for the purpose of comparison. The dotted line shows the statistical model rates (taken from Rauscher \textit{et al.} \cite{Rauscher94AJ}). The shaded band represents the data from Ref.~\cite{CD20N}.}
	\label{fig:rate_gs}
\end{figure}

In Fig.~\ref{fig:rate_gs}, we show the $^{19}$N($n$, $\gamma$)$^{20}$N reaction rates for the g.s. $d_{5/2}$ configuration of $^{20}$N with and without the AMD spectroscopic factors deduced earlier, for the purpose of comparison. Interestingly, the AMD and WS rates are similar till $T_9 \leq$ 1. However, as the temperature rises, the AMD calculations are more consistent with the experimental results. It may be worthwhile to remark that a proper description of $^{20}$N, as a bubble nucleus in its g.s., by the AMD model, was crucial in accounting for the reaction rates. For instance, had the $^{20}$N g.s., built from the g.s. of $^{19}$N$(1/2^-)$ admitted, an $s$-wave neutron that would have destroyed the bubble nature and the consequent reaction rate would have been substantially higher. Incidentally, initial estimates which include coupling the {\it core excited} $^{19}$N$(3/2^-)$ with an $s_{1/2}$ neutron to form the g.s. of $^{20}$N show that the contributions to the reaction rate are possible at higher temperatures~\cite{CD20N}. 

At the same time, the WS reaction rates are almost identical with the AMD and AMD with tail corrected results till T$_{9}$ $\approx$ 1. However, beyond this temperature, discernible differences begin to manifest themselves. It will certainly be exciting to see if this difference is a direct outcome of the extension of the tail of the wave functions at larger distances.  A comparison with statistical rates (taken from Rauscher \textit{et al.} \cite{Rauscher94AJ}) is done through the dotted lines. As is evident, statistical models may not always incorporate the essence of individual nuclear structure of exotic nuclei involved in these capture reactions.

The uncertainty in the g.s. binding energy of 80 keV \cite{NNDC} for $^{20}$N led us to calculate the reaction rates for different one neutron separation energies ranging from 2.08 MeV to 2.24 MeV. We did not see any major difference in the reaction rates throughout this temperature range, and hence, these results are not shown.

We observe that our results from the microscopic as well as phenomenological approach are fairly comparable with the experimental results (shaded data extracted using the experimental cross sections) within their limits of rms statistical uncertainties \cite{CD20N}. 
The overall rate is prone to saturation, as at higher temperatures, other processes like $p$- or $\alpha$-capture begin to interfere and dominate \cite{TSK01AJ}. It will be interesting to further analyse these rates from both the experimental and theoretical view points, especially when the rate for $^{20}$N($n$,$\gamma)^{21}$N is also considered for the abundance of Flourine isotopes via reaction network calculations.

\section{Conclusions}\label{CON}

In this work, we try to bring together various aspects of nuclear physics for a coherent analysis of structure, reaction, and applications to astrophysics. We explore the bubble nature of $^{20}$N, showing its impact on the $^{19}$N($n$,$\gamma)^{20}$N radiative capture reaction rate.

We incorporate the AMD method, a microscopic nuclear structure model, with the fully quantum mechanical theory of finite-range distorted-wave Born approximation (FRDWBA) and analyse the theoretical elastic Coulomb breakup of $^{20}$N on $^{208}$Pb at 256 MeV/u beam energy to give off a $^{19}$N core and a valence neutron. The cross sections generated are used to compute the relevant reaction rates. Such approaches of using a microscopic structure input to the FRDWBA have been successfully used in the past to study halo nuclei~\cite{DCK21EPJ}.

The AMD wave function was corrected for proper asymptotics and was also compared with a phenomenological Woods-Saxon wave function. Although the tail portions may vary, however, the wave function amplitudes are almost similar in the interior region.  With an $S_n$ value of 2.16\,MeV, the nucleus is not very weakly bound and therefore, should have a smaller contribution from tail portions of the wave functions compared to a typical halo nucleus.

The rate of the $^{19}$N($n$,$\gamma)^{20}$N reaction is important for abundance of Fluorine isotopes in stellar nucleosynthesis
and is strongly affected by the underlying structure of $^{20}$N, the bubble structure. The negligibly small spectroscopic factor 
of the $d_{3/2}$ orbit effectively identifies the single particle $d_{5/2}$ state as the sole contributor to the various static properties of $^{20}$N
and the ensuing $^{19}$N($n$,$\gamma)^{20}$N reaction rate.

Thus, a precise knowledge of the structure of exotic nuclei is crucial not only to understand the physics at or near the drip lines, but uncertainties in its description can result in different results for observables like symmetry energy going into the equation of state of infinite neutron-rich nuclear matter \cite{MAC16PRC,CRV09PRL}, which presently is a key to understand the neutron stars and supernovae \cite{Goriely1997,Goriely2013}. It is, therefore, vital that a realistic description of the structure and reactions of exotic nuclei are implemented, which otherwise has to rely on statistical estimates from the available data for more stable nuclei. The FRDWBA can prove to be a handy tool in this regard, given that it requires just the projectile wave function as the input, that can be taken from microscopic structure theories or phenomenological models. This gives us results that match experimental data reasonably well.

\section*{Acknowledgements}
We would like to express our thanks to Prof. Daniel Bemmerer for providing the experimental data sets.
This work was supported by the Scheme for Promotion of Academic and Research Collaboration (SPARC/2018-2019/P309/SL), Ministry of Education (MoE), India and JSPS KAKENHI Grants No. 18K03635, 22H01214, the Collaborative Research Program 2022, Information Initiative Center, Hokkaido University. [VC] also acknowledge a doctoral fellowship form the MoE, India.  [S] acknowledges the European Union's Horizon 2020 research and innovation program under the Marie Skłodowska-Curie Grant Agreement No. 801505. [GS] acknowledges the SID funds 2019 (University of Padova, Italy) under Project No. CASA\_SID19\_01, the European Union’s Horizon 2020 research and innovation program, under Grant Agreement No. 654002, and also the project P20\_01247, funded by the Consejería de Economía, Conocimiento, Empresas y Universidad, Junta de Andalucía (Spain) and ``ERDF-A Way of Making Europe".
	
		
\end{document}